\documentstyle[preprint,aps,epsf]{revtex}

\newcommand{\adots} {{\mathinner{\mkern2mu\raise1pt\hbox{.}\mkern2mu
\raise4pt\hbox{.}\mkern2mu\raise7pt\hbox{.}\mkern1mu}}}

\begin{document}
\title{Renormalization Group in Quantum Mechanics}

\author{Pierre Gosselin}
\address{Universit\'e Grenoble I, Institut Fourier, UMR 5582 CNRS-UJF, \\
UFR de Math\'ematiques, \\
BP74, 38402 Saint Martin d'H\`eres, Cedex, France}

\author{Herv\'e Mohrbach}
\address{LPLI-Institut de Physique, 1 blvd D.Arago, F-57070 Metz, France}

\date{\today}
\maketitle

\begin{abstract}
We establish the renormalization group equation for the running action in
the context of a one quantum particle system. This equation is deduced by
integrating each fourier mode after the other in the path integral
formalism. It is free of the well known pathologies which appear in quantum
field theory due to the sharp cutoff. We show that for an arbitrary
background path the usual local form of the action is not preserved by the
flow. To cure this problem we consider a more general action than usual
which is stable by the renormalization group flow. It allows us to obtain a
new consistent renormalization group equation for the action.
\end{abstract}

\section{\protect\smallskip Introduction}

The renormalization group (RG) \cite{wilson} is a powerful tool for
investigating problems with many degrees of freedom like in quantum field
theory or in the statistical mechanics of phase transitions. Its advantage
is to be non perturbative by nature.\ The RG equation is a flow equation for
an effective action $S_{\Lambda }[\varphi \left( x\right) ]$ under lowering
the ultraviolet cutoff $\Lambda $: 
\[
\Lambda \frac{\partial }{\partial \Lambda }S_{\Lambda }[\varphi ]=\mathcal{F}
[S_{\Lambda }] 
\]
The Wegner Houghton equation \cite{WH} is a kind of RG flow equation in
momentum space where the cutoff $\Lambda $ is the highest momentum of the
field. This equation has been widely studied in the past for problems
related to phase transitions (fixed points, computation of the critical
exponents, etc...).

The RG is not limited to the statistical mechanics or field theory but is
also useful for extracting non-perturbative results in quantum mechanics. In 
\cite{pierre}, we derived the RG equation in momentum space for the
(arbitrary) potential part of a one quantum particle action at zero and
finite temperature (at zero temperature it corresponds to the one
dimensional version of the Wegner Houghton equation). This equation allowed
us to compute, as an example, the ground state and first exited energy level
of the anharmonic oscillator with a great accuracy. The RG equation was
deduced by integrating each mode after the other in the path integral.
Starting with an initial potential $V(x(t))$, we obtain at each step of the
integration an effective potential $V_{m}(x(t)).$ Actually the computation
was done by Taylor expanding the potential $V_{m}(x(t))$ containing a path $
x(t)$ with $m$ modes, around the zero mode path $x_{0}$ (called in this
paper the constant background path in analogy with the constant background
field in field theory) and by integrating only the last mode $x_{m\text{ }}$
in the path integral. In this manner we obtain an equation relating $
V_{m}\left( x_{0}\right) $ and $V_{m-1}\left( x_{0}\right) .$ The
approximation made by projecting on the zero component path $x_{0}$ is named
the Local Potential Approximation (LPA) \cite{morris}. Contrary to the
statistical case where we are looking for the fixed point of the RG
equation, we were interested to integrate this equation until the zero mode (
$m=0$), in order to obtain the so called effective potential $V_{0}\left(
x_{0}\right) $. At zero temperature the effective potential is equal to the
ground state energy \cite{kleinert}, so our procedure was applied to the
computation of the ground state energy of the anharmonic oscillator. This
procedure is summarized in section 2.

\smallskip

It is the aim of the present paper to extend the previous RG\ analysis
beyond the LPA by considering a quantum particle in an arbitrary potential
for an arbitrary background path. This allows us to compute the RG equation
not only for the potential but for the complete action. It is shown that a
local potential is no more consistent because locality is not preserved by
the RG flow. We then choose a particular non local action which is shown to
be preserved along the RG flow but only in the zero temperature limit. The
finite temperature case is then not considered in this paper. In addition it
is shown that the kinetic term with a path independent mass term is kept
constant during the flow (contrary to different papers in field theory),
whereas a path dependant mass term obeys an RG equation too.

\smallskip These results are in contradiction with different works in Field
Theory \cite{branchina} \cite{bonanno}\cite{branchina1}, where the authors
found flow equations for the potential and the kinetic energy for arbitrary
background field preserving the locality assumption. In fact, only the
equation for a constant background field yielding the flow for the potential
is well established. It is well known that the natural extension of this
approximation to non constant background fields obtained by considering a
derivative expansion of the action suffer some pathologies (non analicity)
due to the sharp cutoff used in the computation \cite{morris}\cite{branchina}
\cite{wetterich}. In particular the anomalous dimension can not be computed
with certitude. Then the results obtained in Field theory with a sharp
cutoff must be taken with care. On another hand the introduction of a smooth
cutoff formalism \cite{morris}\cite{wetterich} gives a well defined
mathematical cadre for the computation but the physical quantities depend on
the parameters characterizing the smoothness, the so called scheme
dependence \cite{scheme}.

In the present paper our computation in quantum mechanics (one dimensional
field theory) is free of the just discussed pathologies if we take the limit
of infinite volume (in our case zero temperature) at the end of the
computation, whereas they are present if we start directly in the continuum.
This shows that the generation of non-analytical terms is in fact only an
artifact of the computation in the continuum and disappear if we make a
rigorous discrete calculus.

This paper is organized as follows: we present the general formalism in the
context of the LPA approximation in section 2. In section 3, by\ deriving
the flow of the complete action we find that a new kind of potential must be
necessarily introduced to get a closed form for the action along the flow.
In section 4, we deduce in addition some new equations for the flow of the
coupling constants. The analysis of these flows proves that the LPA is
enough to compute the ground state energy exactly (in the case of a constant
mass term). The renormalization group equation for the kinetic energy is
also computed in section 5, and discussed in view of field theory
applications in section 6.

\smallskip

\section{Renormalization Group Equations of the Potential for a constant
background path}

We work in the euclidean formalism at a finite temperature and discrete time
and quickly retrieve the RG equation for a constant background path as was
done in \cite{pierre}. But here, we limit ourself to the zero temperature
limit, so we will always neglect contributions of order $\frac{1}{\beta ^{2}}
$. Keeping a finite temperature in the intermediary steps allows working
with a finite number of Fourier modes, so that we can integrate each mode
after the other in the path integral.

Consider the euclidean action of a quantum particle at a finite temperature 
\begin{equation}
S(x)=\int_{0}^{\hbar \beta }\left( {\frac{1}{2}}M\left( {\frac{d}{dt}}
x(t)\right) ^{2}+V(x(t))\right) ,
\end{equation}
with $M$ the mass, $V$ the potential. \newline

The effective potential is defined as a constrained path integral over
periodic paths with period $\hbar \beta $
\begin{equation}
\exp \left( -{\beta }V_{0}(x_{0})\right) =\int \mathcal{D}x\delta (\overline{
x}-x_{0})\exp \left( -\frac{1}{\hbar }S(x)\right) \,,
\end{equation}
where $\overline{x}=\frac{1}{\hbar \beta }\int_{0}^{\hbar \beta }dtx(t)$ is
the average position of the particle in the time interval $t\in \left[
0,\beta \right] $.

We consider the Feynman path integral with a discretized time $t_{n}={\frac{
nT}{N+1}}=n\epsilon $, with $N$ an arbitrary large number, and $n=0,\ldots
,N+1$. The Fourier decomposition of a periodic path $x(t_{n})$ contains only
a finite number of Fourier modes 
\begin{equation}
x(t_{n})=x_{0}+{\frac{1}{\sqrt{N+1}}}\sum^{\prime }\exp (i\omega
_{m}t_{n})x_{m}+h.c.,
\end{equation}
where $\sum^{\prime }$ is from $1$ to $\frac{N}{2}$ if $N$ is even and from $
1$ to ${\frac{N-1}{2}}$ if $N$ is odd. The $x_{m}$ are the Fourier modes and 
$\omega _{m}^{2}={\frac{2-cos{\frac{2\pi m}{N+1}}}{\epsilon ^{2}}}$. The
discrete action is: 
\begin{equation}
S_{\frac{N}{2}}(x)=\epsilon \sum_{0}^{\frac{N}{2}}{M}\omega
_{m}^{2}|x_{m}|^{2}+\epsilon \sum_{n=1}^{N+1}V_{\frac{N}{2}}(x(t_{n}))
\end{equation}
The partition function is: 
\begin{equation}
Z=\int {\frac{dx_{0}}{\sqrt{\frac{2\pi \hbar \epsilon }{M}}}}\int \prod_{1}^{
\frac{N}{2}}{\frac{dx_{m}d\bar{x}_{m}}{{\frac{2\pi \epsilon \hbar }{M}}}}
\exp \left( -\frac{1}{\hbar }S_{\frac{N}{2}}\right) \,.
\end{equation}
Now using the fact that $\prod\limits_{1}^{\frac{N}{2}}\epsilon ^{2}\omega
_{m}^{2}=\sqrt{N+1}$ (see \cite{kleinert}) and $\hbar \beta =(N+1)\epsilon $
, we can drop the first integral to get the effective potential 
\begin{equation}
\exp \left( -{\beta }V_{0}(x_{0})\right) =\int \prod_{1}^{\frac{N}{2}}{\frac{
dx_{m}d\bar{x}_{m}}{{\frac{2\pi \epsilon \hbar }{\epsilon ^{2}\omega
_{m}^{2}M}}}}\exp \left( -{\frac{1}{\hbar }}S_{\frac{N}{2}}\right) \,.
\end{equation}
The effective potential is then obtained after integration on all the
Fourier modes except the zero mode, so that it accounts for all the quantum
fluctuations. At zero temperature it is clear that the partition function is
dominated by the minimum of $V_{0}(x_{0})$ which lies at $x_{0}=0$ in the
absence of a phase transition. Then the ground state energy is given by $
E_{0}=V_{0}\left( 0\right) .$ Instead of doing the integration on all the
Fourier modes in one step to compute the effective potential which is in
general a very difficult task and need some approximations, we integrate
only one mode to get an new action with one mode less. This is the spirit of
the RG. Starting with $V_{\frac{N}{2}}(x(t))$ the initial potential we
denote $V_{m}$ the running potential obtained after $N-m$ integrations. To
find the potential $V_{m-1}$ with respect to $V_{m}$ we consider paths with
only one Fourier mode: 
\begin{equation}
x(t_{n})=x_{0}+{\frac{1}{\sqrt{N+1}}}exp(i\omega _{m}t_{n})x_{m}+h.c.
\label{path}
\end{equation}
By integrating on the mode $x_{m}$ we will obtain a new potential $V_{m-1}$
which could depend only on the zero mode $x_{0}$ due to the particular path
chosen. We could as well choose a different path than (\ref{path}). Then the
potential $V_{m-1}$would be a function of another path. The particular path
chosen is called the background path, and for this reason $x_{0}$ is called
the constant background path.

Then by definition we have the relation: 
\begin{eqnarray}
{\exp \left( -\beta V_{m-1}(x_{0})\right) \hfill } &=&{\hfill \int {\frac{
dx_{m}d\bar{x}_{m}}{{\frac{2\pi \epsilon \hbar }{\epsilon ^{2}\omega
_{m}^{2}M}}}}\exp {\Huge (}}-{{\frac{\epsilon }{\hbar }{\LARGE (}M}\omega
_{m}^{2}|x_{m}|^{2}+}  \nonumber \\
&&\sum_{n=0}^{N+1}V_{m}(x_{0}+{\frac{exp(i\omega _{m}t_{n})x_{m}}{\sqrt{N+1}}
}+h.c.){\LARGE )}{\Huge )}{.}
\end{eqnarray}
Expanding the potential $V_{m}$ to second order around the point $x_{0}$ and
summing over $n$, yields to: 
\[
{\exp \left( -\beta V_{m-1}(x_{0})\right) =\exp \left( -\beta
V_{m}(x_{0})\right) \int {\frac{dx_{m}d\bar{x}_{m}}{{\frac{2\pi \epsilon
\hbar }{\epsilon ^{2}\omega _{m}^{2}M}}}}\exp {\Huge (}}-{{\frac{\epsilon }{
\hbar }{\LARGE (}M}\omega _{m}^{2}+}V_{m}^{(2)}(x_{0}){\LARGE )}{|x_{m}|^{2}}
+O(\frac{1}{\beta ^{2}}){\Huge )} 
\]
After the gaussian integration on the modes $x_{m}$ and $\overline{x}_{m}$
we find the following RG equation for the potential 
\begin{equation}
V_{m-1}(x_{0})=V_{m}(x_{0})+\frac{1}{\beta }\log (1+\frac{V_{m}^{(2)}(x_{0})
}{M\omega _{m}^{2}})+O(\frac{1}{\beta ^{2}})  \label{potfx0}
\end{equation}
where $V_{m}^{(2)}(x_{0})=\frac{d^{2}}{dx_{0}^{2}}V_{m}(x_{0})$.

This equation is the one dimensional version of the Wegner-Houghton equation
in the LPA. It is this equation which has been studied in the past in
statistical mechanics of phase transition. Going beyond this approximation
is notorious difficult as discussed in the introduction. We will go back to
this question in the last section. Only in the D=1 case is the computation
out of trouble as shown in the following.

\smallskip

As discussed in \cite{pierre}, it is very difficult to integrate numerically
the non-linear equation (\ref{potfx0}) in order to find the ground state
energy. It's much more easier to do a Taylor expansion of the running
potential in order to compute the flow of the coupling constants.\
Unfortunately we have to truncate this expansion and the results could not
be exact, but are nevertheless impressively good (see \cite{pierre}).
Defining the $n$th coupling constant at the scale $m$ by $g_{m}^{(n)}(x_{0})=
\frac{d^{n}}{dx_{0}^{n}}V_{m}(x_{0})$ we can get the flow of all the
coupling constants by deriving (\ref{potfx0}). Then the Taylor expansion of
the potential $V_{m}(x(t_{n}))$ around $x_{0}$ yields to the following
expansion of the interaction in the Fourier space at the cutoff scale $m$: 
\begin{eqnarray}
\frac{\epsilon }{\beta }\sum_{n=1}^{N+1}V_{m}(x(t_{n}))
&=&g_{m}^{(0)}(x_{0})+\frac{g_{m}^{(2)}(x_{0})}{2!(N+1)}
\sum_{p=-m}^{m}x_{p}x_{-p}+  \nonumber \\
&&\frac{g_{m}^{(4)}(x_{0})}{4!(N+1)^{2}}\sum_{p_{1}...p_{4}}^{m}\delta
_{p_{1}+...+p_{4},0}x_{p_{1}}x_{p_{2}}x_{p_{3}}x_{p_{4}}+...  \label{potexp}
\end{eqnarray}
where $g_{m}^{(0)}(x_{0})=V_{m}(x_{0})$. It will be shown below that this
expansion is actually not preserved by the renormalization group for a non
constant background path. A more general potential still local in time must
be introduced to avoid some inconsistencies which appear when one tries to
deduce the RG equation for an non constant background path.

\section{Renormalization Group Equations of the Potential for an arbitrary
background path}

In this section we keep the background path arbitrary. Instead of computing $
V_{m-1}(x_{0})$ we want to deduce the running action $S_{m-1}(x^{(m-1)})$
from the action $S_{m}(x^{(m)})$ where we define 
\begin{equation}
x^{(m)}(t_{n})=x_{0}+{\frac{1}{\sqrt{N+1}}}\sum_{p=1}^{m}\exp (i\omega
_{p}t_{n})x_{p}+h.c.,
\end{equation}
the truncated path with Fourier components up to $m$.

Analogously to the preceding section one defines an action at step $m-1$ by
integrating only on the two modes $x_{m}$ and $\overline{x}_{m}$ that is: 
\begin{equation}
\exp \left( -{\frac{1}{\hbar }}S_{m-1}(x^{(m-1)})\right) =\int {\frac{dx_{m}d
\bar{x}_{m}}{{\frac{2\pi \epsilon \hbar }{\epsilon ^{2}\omega _{m}^{2}M}}}}
\exp \left( -\frac{1}{\hbar }S_{m}(x^{(m)})\right) \,.
\end{equation}
To get only the contributions of order $\frac{1}{\beta }$ it is enough to
expand the action $S_{m}(x^{(m)})$ around $x^{(m-1)}$ to the second order 
\cite{pierre}, so that the result is obtained after a gaussian integration: 
\begin{equation}
S_{m-1}(x^{(m-1)})=S_{m}(x^{(m-1)})+{\frac{1}{2}}\log (det(\mathcal{A}
))-J^{t}\mathcal{A}^{-1}J\,
\end{equation}
with $\mathcal{A}$ the symmetric $2\times 2$ matrix
 
\begin{eqnarray}
\mathcal{A}=
\pmatrix{1+\sum_{n=0}^{N+1}\frac{V_{m}^{(2)}(x^{(m-1)}(t_{n}))(1+\cos (2\omega
_{m}t_{n}))}{(N+1)\omega _{m}^{2}M}  && -\sum_{n=0}^{N+1}\frac{
V_{m}^{(2)}(x^{(m-1)}(t_{n}))\sin (2\omega _{m}t_{n})}{(N+1)\omega _{m}^{2}M}\cr
-\sum_{n=0}^{N+1}\frac{V_{m}^{(2)}(x^{(m-1)}(t_{n}))\sin (2\omega
_{m}t_{n})}{(N+1)\omega _{m}^{2}M} && 1+\sum_{n=0}^{N+1}\frac{
V_{m}^{(2)}(x^{(m-1)}(t_{n}))(1-\cos (2\omega _{m}t_{n}))}{(N+1)\omega
_{m}^{2}M} \cr} \nonumber
\end{eqnarray}

and 
\begin{eqnarray}
J=\pmatrix{
\frac{\epsilon }{\sqrt{N+1}}\sum_{n=0}^{N+1}V_{m}^{(1)}(x^{(m-1)}(t_{n}))
\cos (\omega _{m}t_{n}) \cr 
-\frac{\epsilon }{\sqrt{N+1}}
\sum_{n=0}^{N+1}V_{m}^{(1)}(x^{(m-1)}(t_{n}))\sin (\omega _{m}t_{n})\cr}
\nonumber
\end{eqnarray}

Now we consider a mass term which is position independent. As we shall see a
posteriori, this term will stay constant during the RG flow (a priori there
was a possibility to get a flowing mass $M_{m}$ as well as generation of
higher derivative interactions). Then we write only the equation for the
running potential

\begin{eqnarray}
\epsilon \sum_{n=0}^{N+1}V_{m-1}(x^{(m-1)}(t_{n})) &=&\epsilon
\sum_{n=0}^{N+1}V_{m}(x^{(m-1)}(t_{n}))+{\frac{1}{2}}\log \left( \left(
1+\sum_{n=0}^{N+1}{\frac{V_{m}^{(2)}(x^{(m-1)}(t_{n}))}{(N+1)\omega _{m}^{2}M
}}\right) ^{2}\right.  \nonumber \\
&&\left. -\sum_{n_{1}=0}^{N+1}\sum_{n_{2}=0}^{N+1}{\frac{
V_{m}^{(2)}(x^{(m-1)}(t_{n_{1}}))}{(N+1)\omega _{m}^{2}M}}{\frac{
V_{m}^{(2)}(x^{(m-1)}(t_{n_{2}}))}{(N+1)\omega _{m}^{2}M}}\cos (2\omega
_{m}(t_{n_{2}}-t_{n_{1}}))\right)  \nonumber \\
&&-J^{t}A^{-1}J.  \label{equa}
\end{eqnarray}
Expanding $V_{m}^{(2)}$ around $x_{0}$ in the left hand side of (\ref{equa})
doesn't generate a flow for the kinetic term, so the logarithm term
contributes only to the potential (we can check that this is also true for
the 'source 'term). The explicit form of $J^{t}A^{-1}J$ is not necessary for
the point we want to show. The problem is the appearance of the non local
double sum in the logarithm 
\begin{equation}
\sum_{n_{1}=0}^{N+1}\sum_{n_{2}=0}^{N+1}{\frac{
V_{m}^{(2)}(x^{(m-1)}(t_{n_{1}}))}{(N+1)\omega _{m}^{2}M}}{\frac{
V_{m}^{(2)}(x^{(m-1)}(t_{n_{2}}))}{(N+1)\omega _{m}^{2}M}}\cos (2\omega
_{m}(t_{n_{2}}-t_{n_{1}}))
\end{equation}
which makes us impossible to obtain a local potential for $V_{m-1}$. That
is, if we have a local potential of the form 
\begin{equation}
V_{m}(x^{(m)})=V_{m}(x_{0}+{\frac{1}{\sqrt{N+1}}}\sum_{p=1}^{m}\exp (i\omega
_{p}t_{n})x_{p}+h.c.),
\end{equation}
the potential at the scale $m-1$ cannot be written: 
\begin{equation}
V_{m-1}(x^{(m-1)})=V_{m-1}(x_{0}+{\frac{1}{\sqrt{N+1}}}\sum_{p=1}^{m-1}\exp
(i\omega _{p}t_{n})x_{p}+h.c.)
\end{equation}
In other words the assumption concerning the expansion of the running
potential (\ref{potexp}) is too strong and this kind of expansion is not
preserved by the RG. An other way to see this problem is to replace $V_{m}$
in the r.h.s of (\ref{equa}) by it's expansion (\ref{potexp}). It's then
easy to check that the expansion obtained for $V_{m-1}$ is of a different
kind than (\ref{potexp}). Actually more coupling constants are introduced.
We deduce that the usual expansion (\ref{potexp}) is not preserved along the
RG flow.

A consequence of this unclosedness can easily be seen by developing the
Renormalization group equation (\ref{equa}) around $x_{0}$ to compute the
equation for the second derivative of the potential at the point $x_{0}$.
The zero order of this expansion gives the RG equation for the running
potential 
\[
V_{m-1}(x_{0})=V_{m}(x_{0})+{\frac{1}{\beta }}\log \left( 1+{\frac{
V_{m}^{(2)}(x_{0})}{\omega _{m}^{2}M}}\right) , 
\]
which is the usual one (\ref{potfx0}), whereas the second order gives

\begin{equation}
V_{m-1}^{(2)}(x_{0})=V_{m}^{(2)}(x_{0})+{\frac{1}{\beta }}{\frac{
V_{m}^{(4)}(x_{0})}{\omega _{m}^{2}M+V_{m}^{(2)}(x_{0})}}.  \label{potx2p}
\end{equation}
These two equations are evidently inconsistent with each other. This is
again a manifestation of the fact that the local ansatz for the potential at
the scale $m$ is not preserved after one step of the RG, and a kind of non
locality has to be introduced.

\section{Generalized potential}

To avoid the just mentioned inconsistency, we have to consider a class of
potentials which is preserved by the renormalization group flow. Looking to
the RG equation leads us to consider for the potential $V_{m}$ a function of
the $m+1$ independent variables $V_{m}(x_{0},\ldots ,x_{p}e^{i\omega
_{p}t}+x_{-p}e^{i\omega _{p}t},\ldots ,e^{i\omega _{m}t}x_{m}+e^{-i\omega
_{m}t}x_{-m})$ denoted again for convenience by $V_{m}(x^{(m)}(t))$.

In fact, due to the non linearity of the $\omega _{m}$, this kind of
potential is not preserved by the RG flow. But recall that we want to
consider the limit $\beta \to \infty $. In such limit $\omega _{m}=\frac{
2\pi m}{\hbar \beta }$, and we will show that the class of function
considered is preserved by the flow.

To derive the new flow equation for the potential, it's easier to work in
functional space. We write the action at scale $m$
\begin{equation}
S_{m}(x^{(m)})=\epsilon \sum_{p=0}^{m}{M}\omega _{p}^{2}|x_{p}|^{2}+\beta
U_{m}(x^{(m)})
\end{equation}
where we have introduced the notation 
\begin{equation}
U_{m}(x^{(m)})=\frac{\epsilon }{\beta }\sum_{n=0}^{N+1}V_{m}(x_{0},e^{i
\omega _{1}t}x_{1}+e^{-i\omega _{1}t}x_{-1},...,e^{i\omega
_{m}t}x_{m}+e^{-i\omega _{m}t}x_{-m})  \label{umpo1}
\end{equation}
Expanding the potential around $x_{0}$ we get the following expansion: 
\begin{equation}
U_{m}(x^{(m)})=g_{m}^{0}(x_{0})+\sum_{n=-m}^{m}\frac{
g_{m}^{n,-n}(x_{0})}{2!(N+1)}x_{n}x_{-n}+\sum_{n_{1},...,n_{4}=-m}
^{m}\frac{g_{m}^{n_{1}...n_{4}}(x_{0})}{4!(N+1)^{2}}
x_{n_{1}}x_{n_{2}}x_{n_{3}}x_{n_{4}}\delta _{n_{1}+...+n_{4},0}+...
\label{umf1}
\end{equation}
\smallskip which contains much more coupling constants than equation (\ref
{potexp}): as usual each coupling constant is cutoff dependent but acquires
now in addition a dependence in the Fourier modes of the field. Then, most
of these couplings disappear when the cutoff $m$ reaches the value of its
largest momentum. Note that in (\ref{umf1}) there is still conservation of
the momentum due to our choice of the potential (\ref{umpo1}).

The RG equation is still: 
\begin{equation}
S_{m-1}(x^{(m-1)})=S_{m}(x^{(m-1)})+{\frac{1}{2}}\log (det(\mathcal{A}
))-J^{t}\mathcal{A}^{-1}J\,,  \label{graction1}
\end{equation}
but the matrix $\mathcal{A}$ is now
 
\begin{eqnarray}
\mathcal{A}=\pmatrix{
1+\frac{U_{m}^{(m,-m)}+Re(U_{m}^{(m,m)})}{M\omega _{m}^{2}} & -\frac{
Im(U_{m}^{(m,m)})}{M\omega _{m}^{2}} \cr 
-\frac{Im(U_{m}^{(m,m)})}{M\omega _{m}^{2}} & 1+\frac{U_{m}^{(m,-m)}-
Re(U_{m}^{(m,m)})}{M\omega _{m}^{2}} \cr}
\end{eqnarray}

and 
\begin{eqnarray}
J=\pmatrix{ Re(U_{m}^{(m)}) \cr 
-Im(U_{m}^{(m)}) \cr}
\end{eqnarray}

with the notation: 
\begin{equation}
U_{m}^{(n_{1}...n_{p})}\equiv (N+1)^{\frac{p}{2}}\frac{\partial ^{p}U_{m}}{
\partial x_{n_{1}}...\partial x_{n_{p}}}\left|
_{(x_{n_{1}},...,x_{n_{p}})=0}\right.  \label{derivU1}
\end{equation}
The flow equation for the potential is now:

\begin{equation}
U_{m-1}=U_{m}+\frac{1}{2\beta }\lg {\LARGE (}(1+\frac{U_{m}^{(m,-m)}}{
M\omega _{m}^{2}})^{2}-\left| \frac{U_{m}^{(m,m)}}{M\omega _{m}^{2}}\right|
^{2}){\LARGE )}
-\frac{(M\omega _{m}^{2}+U_{m}^{(m,-m)})\left|
U_{m}^{(m)}\right| ^{2}-Re(U_{m}^{(-m,-m)}U_{m}^{(m)2})}{(M\omega
_{m}^{2}+U_{m}^{(m,-m)})^{2}-\left| U_{m}^{(m,m)}\right| ^{2}}  \label{potx1}
\end{equation}

The last contribution in the right hand side comes from the 'source term'.
In this expression $U_{m}$ and its various derivatives are functions of the
path $x^{(m-1)}$. It's easy to check that the Taylor expansion of $U_{m-1}$
is of the form (\ref{umf1}). In addition, the r.h.s of (\ref{potx1}) does
not give any contributions to the kinetic term.

We deduce the equation for a constant background path $x_{0}$ which now
reads:

\begin{equation}
U_{m-1}(x_{0})=U_{m}(x_{0})+{\frac{1}{\beta }}\log \left( 1+{\frac{{U}
_{m}^{(m,-m)}(x_{0})}{M\omega _{m}^{2}}}\right) .  \label{pottnouv1}
\end{equation}
Expanding (\ref{potx1}) to the second order in $x_{0}$ allows us to write
for the second derivative of the potential:

\begin{equation}
{U}_{m-1}^{(p,-p)}(x_{0})=U_{m}^{(p,-p)}(x_{0})+{\frac{1}{\beta }}{\frac{{U}
_{m}^{(p,-p,m,-m)}(x_{0})}{M\omega _{m}^{2}+{U}_{m}^{(m,-m)}(x_{0})}}.
\label{pott2nouv1}
\end{equation}
In this formula the meaning of derivative is similar to (\ref{derivU1}),
that is: 
\begin{equation}
{U}_{m-1}^{(p,-p)}(x_{0})=(N+1)\frac{\partial ^{2}U_{m}}{\partial
x_{p}\partial x_{-p}}\left| _{(x^{(m-1)})=x_{0}}\right.
\end{equation}
Equations (\ref{pottnouv1}) and (\ref{pott2nouv1}) are no more inconsistent
: one cannot obtain the second equation by deriving the first one because
all the variables except $x_{0}$ are already set to zero.

Note that the RG equation (\ref{pottnouv1}) for the potential is now
different from (\ref{potfx0}) so new equations for the running coupling
constants are expected.

\section{Flow equations of the coupling constants}

This time we choose to expand $U$ around $x_{0}=0$. The $n$th coupling
constant at the scale $m$ is defined as 
\begin{equation}
\delta _{n_{1}+...+n_{p},0}g_{m}^{n_{1},\ldots ,n_{p}}=(N+1)^{\frac{p}{2}}
\frac{\partial ^{p}U_{m}}{\partial x_{n_{1}}...\partial x_{n_{p}}}\left|
_{0}\right.
\end{equation}
so that:

\begin{equation}
U_{m}(x^{(m)})=\sum_{p=0}^{\infty }\sum_{n_{1},...,n_{p}=-m}^{m}\frac{
g_{m}^{n_{1},\ldots ,n_{p}}}{p!(N+1)^{\frac{p}{2}}}x_{n_{1}}\ldots
x_{n_{p}}\delta _{n_{1}+...+n_{p},0}
\end{equation}
The equation for the first coupling constant is: 
\begin{equation}
g_{m-1}^{0}=g_{m}^{0}+\frac{1}{\beta }\log (1+\frac{g_{m}^{m,-m}}{M\omega
_{m}^{2}})
\end{equation}
The value of the ground state energy is given by $E_{0}=g_{0}^{0}=V_{0}(0)$
which is the minimum of the effective potential. The flows of the quadratic
coupling constants for $p\leq m-1$ are

\begin{equation}
g_{m-1}^{p,-p}=g_{m}^{p,-p}+\frac{1}{\beta }\frac{g_{m}^{m,-m,p,-p}}{M\omega
_{m}^{2}+g_{m}^{m,-m}}.
\end{equation}
The particular value $g_{0}^{0,0}$ corresponds to the mass gap or the
inverse correlation length in statistical mechanics language. Then it is
well known that the first exited energy level can be deduced from the
relation $E_{1}-E_{0}=\sqrt{g_{0}^{0,0}}$ .

The flows of the four and six order couplings are given below

\begin{equation}
g_{m-1}^{p_{1},p_{2},p_{3},p_{4}}=g_{m}^{p_{1},p_{2},p_{3},p_{4}}+\frac{1}{
\beta }{\Large (}{\frac{g_{m}^{p_{1},p_{2},p_{3},p_{4},m,-m}}{M\omega
_{m}^{2}+g_{m}^{m,-m}}}-{\frac{
g_{m}^{p_{1},p_{2},m,-m}g_{m}^{p_{3},p_{4},m,-m}{+Perm(}
p_{1},p_{2},p_{3},p_{4})}{(M\omega _{m}^{2}+g_{m}^{m,-m})^{2}}}{\Large ),}
\end{equation}
\begin{eqnarray}
g_{m-1}^{p_{1},p_{2},p_{3},p_{4},p_{5},p_{6}}
&=&g_{m}^{p_{1},p_{2},p_{3},p_{4},p_{5},p_{6}}+\frac{1}{\beta }{\Large (}
\frac{g_{m}^{p_{1},p_{2},p_{3},p_{4},p_{5},p_{6},m,-m}}{(M\omega
_{m}^{2}+g_{m}^{m,-m})}{-}\frac{
g_{m}^{p_{1},p_{2},m,-m}g_{m}^{p_{3},p_{4},p_{5},p_{6},m,-m}+Perm}{(M\omega
_{m}^{2}+g_{m}^{m,-m})^{2}}  \nonumber \\
&&+\frac{
g_{m}^{p_{1},p_{2},m,-m}g_{m}^{p_{3},p_{4},m,-m}g_{m}^{p_{5},p_{6},m,-m}+Perm
}{(M\omega _{m}^{2}+g_{m}^{m,-m})^{3}}{\Large )}  \nonumber \\
&&{\Large -}{\frac{
g_{m}^{p_{1},p_{2},p_{3},m}g_{m}^{p_{4},p_{5},p_{6},-m}+Perm}{M\omega
_{m}^{2}+g_{m}^{m,-m}}}
\end{eqnarray}
where the last term is the tree level contribution of the ''source term''.
In these relations the conservation of the momenta are implicitly supposed
as well as the condition that all the momenta are smaller than the cutoff.
Apparently we see a different flow for each coupling constant. Some of them
disappear when the cutoff reaches the largest momentum of the coupling. A
tree level (source term) renormalization appears for the couplings bigger
than the fourth order.

Suppose we want to compute the ground state energy of the anharmonic
oscillator whose potential is: 
\begin{equation}
V_{\frac{N}{2}}(x)=\frac{M\Omega ^{2}}{2}x^{2}+\frac{\lambda }{4!}x^{4}
\end{equation}
For this initial potential, the coupling constants are Fourier modes
independent. It is straightforward that the running coupling constants split
into different families. For example $g_{m}^{p,-p},$ $g_{m}^{p,-p,q,-q},$ $
g_{m}^{p,-p,q,-q,r,-r}$ etc, are momentum independent. We can thus introduce
the notation 
\begin{eqnarray}
g_{m}^{p,-p} &=&g_{m}^{(2)}  \nonumber \\
g_{m}^{p,-p,q,-q} &=&g_{m}^{(4)}  \nonumber \\
g_{m}^{p,-p,q,-q,r,-r} &=&g_{m}^{(6)}\text{ \qquad etc...}  \label{coup1}
\end{eqnarray}
To compute the ground state and the first exited state energy as done in 
\cite{pierre}, one can see that only the kind of coupling constants in (\ref
{coup1}) are necessary. This explain the great accuracy of the computation
based on equation (\ref{potfx0}): the LPA gives precisely the flow of the
coupling constants allowing for the computation of the ground state energy.
The other coupling constants follow different flows without influencing the
preceding ones.

\section{Renormalization of the kinetic term}

In the preceding sections we have shown that the kinetic energy is RG
invariant, that a mass term chosen position independent is constant along
the flow. In this section we compute the flow equation of a generalized
kinetic energy term. The corresponding action written in the continuum is
now: 
\begin{equation}
S(x)=\int_{0}^{\hbar \beta }\left( {\frac{1}{2}}Z(x(t))\left( {\frac{d}{dt}}
x(t)\right) ^{2}+V(x(t))\right)
\end{equation}
At scale $m$ we make the same ansatz for the potential energy as before. We
write the kinetic term in the Fourier space as: 
\begin{eqnarray*}
&&\frac{\epsilon }{\beta }\sum_{n=0}^{N+1}Z_{m}(x_{0},e^{i\omega
_{1}t_{n}}x_{1}+e^{-i\omega _{1}t_{n}}x_{-1},...,e^{i\omega
_{m}t_{n}}x_{m}+e^{-i\omega _{m}t_{n}}x_{-m})e^{i(\omega _{i}+\omega
_{j})t_{n}}\omega ^{i}\omega ^{j}x_{i}x_{j} \\
&=&-Z_{m,i+j}(x^{(m)})\omega ^{i}x^{i}\omega ^{j}x^{j}
\end{eqnarray*}
where the Fourier coefficient $Z_{m,i+j}$ of $Z_{m}(x^{(m)})$ is defined by 
\begin{equation}
Z_{m,i+j}=\frac{\epsilon }{\beta }\sum_{n=0}^{N+1}Z_{m}e^{i(\omega
_{i}+\omega _{j})t_{n}}.
\end{equation}
In order to follow as close as possible the computation of the preceding
section we introduce the notation for the coefficient $\omega
_{m}^{2}x_{m}x_{-m}$: 
\begin{equation}
M_{m}\equiv Z_{m,0}(x^{(m-1)}).
\end{equation}
The RG equation for the discretized action is then again: 
\begin{equation}
\exp \left( -{\frac{1}{\hbar }}S_{m-1}(x^{(m-1)})\right) =\int \frac{dx_{m}d
\bar{x}_{m}}{\frac{2\pi \hbar }{\epsilon \omega _{m}M_{m}}}\exp \left( -
\frac{1}{\hbar }S_{m}(x^{(m)})\right) \,.
\end{equation}
Again we compute the path integral by expanding the kinetic term to get all
the terms quadratic in $x_{m}$ and compute the gaussian integral. The
following conditions between the Fourier transform of the derivatives of the
mass term are needed: 

\begin{eqnarray}
Z_{m,k}^{(m)}&=&Z_{m,k}^{(-m)} \nonumber \\ 
Z_{m,k}^{(m)} &&\text{ is real} \nonumber \\ 
Z_{m,k}^{(m,m)}&=&\overline{Z_{m,k}^{(-m,-m)}} \nonumber \\ 
Z_{m,-2m-i}^{(m)}&=&-Z_{m,2m-i}^{(m)} \nonumber
\end{eqnarray}

The RG equation is still given by (\ref{graction1}) with the following
matrices: 
\begin{eqnarray}
\mathcal{A}=\pmatrix{
1+\frac{A_{m}^{(m,-m)}+Re(B_{m}^{(m,m)})}{M_{m}\omega _{m}^{2}} & -
\frac{Im(B_{m}^{(m,m)})}{M_{m}\omega _{m}^{2}} \cr 
-\frac{Im(B_{m}^{(m,m)})}{M_{m}\omega _{m}^{2}} & 1+\frac{
A_{m}^{(m,-m)}-Re(B_{m}^{(m,m)})}{M_{m}\omega _{m}^{2}} \cr}
\end{eqnarray}

\begin{eqnarray}
J=\pmatrix{Re(C_{m}^{(m)}) \cr 
-Im(C_{m}^{(m)}) \cr}
\end{eqnarray}

where 
\begin{equation}
A_{m}^{(m,-m)}=U_{m}^{(m,-m)}+Z_{m,i+j}^{(m,-m)}\omega ^{i}x^{i}\omega
^{j}x^{j}
\end{equation}
and 
\begin{equation}
B_{m}^{(m,m)}=U_{m}^{(m,m)}+Z_{m,2m}\omega
_{m}^{2}+Z_{m,2m+i+j}^{(m,m)}\omega ^{i}x^{i}\omega ^{j}x^{j}-\partial
_{j}Z_{m,2m+i+j}^{(m)}\omega ^{i}x^{i}\omega
^{j}x^{j}-Z_{m,2m+i}^{(m)}(\omega ^{i})^{2}x^{i}
\end{equation}
and 
\begin{equation}
C_{m}^{(m)}=U_{m}^{(m)}-Z_{m,m+i}(\omega ^{i})^{2}x^{i}-\partial
_{j}Z_{m,m+i+j}\omega ^{i}x^{i}\omega ^{j}x^{j}+Z_{m,m+i+j}^{(m)}\omega
^{i}x^{i}\omega ^{j}x^{j}
\end{equation}
where all the terms $A,B,C,U,Z$ are functions of the path $x^{(m-1)}$, and
obviously the indices $i,j$ are different from $m$ or $-m$. The notation $
\partial _{j}Z$ means that $x_{j}\neq 0$ as opposed to the notation $Z^{(j)}$
.

With these notations the renormalization group equation for the complete
action reads:

\begin{equation}
S_{m-1}=S_{m}+\frac{1}{2\beta }\lg ((1+\frac{A_{m}^{(m,-m)}}{M_{m}\omega
_{m}^{2}})^{2}-\left| \frac{B_{m}^{(m,m)}}{M_{m}\omega _{m}^{2}}\right| ^{2}))
-\frac{(M\omega _{m}^{2}+A_{m}^{(m,-m)})\left| C_{m}^{(m)}\right|
^{2}-Re(B_{m}^{(-m,-m)}C_{m}^{(m)2})}{(M_{m}\omega
_{m}^{2}+A_{m}^{(m,-m)})^{2}-\left| B_{m}^{(m,m)}\right| ^{2}}  \label{rgs1}
\end{equation}
We can now deduce the equation for the potential and the kinetic term. The
contributions to the potential come from terms which are $\omega
^{i}x^{i}\omega ^{j}x^{j}$ independent. The flow equation of the mass term
is found by keeping the contributions of the form $\omega ^{i}x^{i}\omega
^{j}x^{j}$ and neglecting higher order derivatives interactions which are
now generated by the RG. Remember that for a constant mass term there isn't
any flow of the kinetic energy.

The RG equation for the kinetic term is: 
\begin{eqnarray}
Z_{m-1,i+j} &=&Z_{m,i+j}+\frac{1}{2\beta }\text{{\Huge [}}\frac{
Z_{m,i+j}^{(m,-m)}{\LARGE (}M_{m}\omega _{m}^{2}+U_{m}^{(m,-m)}{\LARGE )}}{G}
-  \nonumber \\
&&\text{{\Huge (}}\frac{(Z_{m,i+j-2m}^{(-m,-m)}-\partial
_{j}Z_{m,i+j-2m}^{(m)})(U_{m}^{(m,m)}+Z_{m,2m}\omega _{m}^{2})}{G}  \nonumber
\\
&&-\frac{(\partial _{j}Z_{m,2m}\omega _{m}^{2}+\partial
_{j}U_{m}^{(m,m)})Z_{m,i-2m}^{(m)}+(Z_{m,2m}\omega
_{m}^{2}+U_{m}^{(m,m)})\partial _{j}Z_{m,i-2m}^{(m)}}{G}  \nonumber \\
&&+Perm(m\leftrightarrow -m)\text{{\Huge )]}}+{\LARGE (}\text{contribution
of the source term}{\LARGE )}  \label{equazkl}
\end{eqnarray}
where 
\begin{equation}
G=(M_{m}\omega _{m}^{2}+U_{m}^{(m,-m)})^{2}-\left|
U_{m}^{(m,m)}+Z_{m,2m}\omega _{m}^{2}\right| ^{2}
\end{equation}
The flow of the potential is now:

\begin{equation}
U_{m-1}=U_{m}+\frac{1}{2\beta }\lg \left( (1+\frac{U_{m}^{(m,-m)}}{
M_{m}\omega _{m}^{2}})^{2}-\left| \frac{U_{m}^{(m,m)}}{M_{m}\omega _{m}^{2}}
\right| ^{2}\right) {\LARGE +(}\text{contribution of the source term}{\LARGE 
).}
\end{equation}
The source term is quite complicate and not very illuminating, this is why
we didn't write it. Rather, by expanding the equations around $x_{0}$, we
get the couple of equations concerning the potential and the kinetic term
for a constant background field. In that case the source terms are zero. We
get only the flow of the zero component Fourier transform of $Z$ noted $
M_{m}\equiv Z_{m}(x_{0})$. This equation is deduced by considering (\ref
{equazkl}) for $i+j=0$, 
\begin{equation}
Z_{m-1}(x_{0})=Z_{m}(x_{0})+\frac{1}{2\beta }{\Huge [}\frac{
Z_{m}^{(m,-m)}(x_{0})}{Z_{m}(x_{0})\omega _{m}^{2}+U_{m}^{(m,-m)}(x_{0})}
{\Huge ]\;.}
\end{equation}
Naturally we recover the equation for the effective potential: 
\begin{equation}
U_{m-1}(x_{0})=U_{m}(x_{0})+{\frac{1}{\beta }}\log \left( 1+\frac{{U}
_{m}^{(m,-m)}(x_{0})}{Z_{m}(x_{0})\omega _{m}^{2}}\right) .
\label{potefflast1}
\end{equation}
Now recall that for large N, $\omega _{m}^{2}$ is equivalent to $(\frac{2\pi
m}{\hbar \beta })^{2}$ and then runs from $0$ to $(\frac{\pi }{\epsilon }
)^{2}$ . Rewriting (\ref{potefflast1}) as 
\[
U_{m-1}(x_{0})=U_{m}(x_{0})+\frac{\hbar }{2\pi }\frac{\pi }{\epsilon }\frac{2
}{N+1}\log \left( 1+\frac{{U}_{m}^{(m,-m)}(x_{0})}{Z_{m}(x_{0})\left( \frac{
\pi }{\epsilon }\frac{2m}{N+1}\right) ^{2}}\right) , 
\]
the limit $N\rightarrow \infty $ (zero temperature) gives a continuous
equation \cite{pierre} 
\begin{equation}
U_{k-\Delta k}(x_{0})=U_{k}(x_{0})+\frac{\hbar \Delta k}{2\pi }\log \left( 1+
\frac{{U}_{k}^{(k,-k)}(x_{0})}{Z_{k}(x_{0})k^{2}}\right) ,
\label{potcontinue1}
\end{equation}
where we have introduced the notations $k^{2}=\omega _{m}^{2}$, $\Delta k=
\frac{2\pi }{\epsilon (N+1)}$ . In the same manner we get a continuous
equation for $Z_{k}(x_{0})$ which is: 
\begin{equation}
Z_{k-\Delta k}(x_{0})=Z_{k}(x_{0})+\frac{\hbar \Delta k}{4\pi }{\Huge [}
\frac{Z_{k}^{(k,-k)}(x_{0})}{(Z_{k}(x_{0})k^{2}+U_{k}^{(k,-k)}(x_{0})}{\Huge 
]\;.}  \label{eqZ0c1}
\end{equation}
From (\ref{eqZ0c1}) we see that if in the initial Lagrangian the mass term
is constant $(Z_{\frac{N}{2}}^{(\frac{N}{2},-\frac{N}{2})}=0)$ it will stay
constant during the RG procedure as found in section 2, and no higher
derivative interactions will be generated (only the potential changes). This
result is in contradiction with most of the results found in field theory.
We discuss this point in the next section.

Note also that, by keeping a finite volume (finite temperature) we were able
to integrate one mode after the other. The continuous differential equations
were derived by taking the infinite volume limit only at the end of the
computation. In the next section we derive again the RG equation by
considering the infinite volume (zero temperature) from the beginning to
show that non-analytical terms appear in this case.

\section{Generalization to Field Theory}

The point is the generation of non-analytic terms in Field Theory. These
terms appear when the integration on the fast modes is performed on a
spherical shell of thinness $\Delta k$. In fact doing again our computation
in quantum mechanics directly in the continuum, we face exactly the same
problem as in Field Theory. To simplify the computation, suppose that the
background field is made of a constant plus one mode configuration. Let's
try to compute the RG equation for $Z_{k}$ and $U_{k}$. For the point we
want to show, we don't need to consider the generalized potential and it's
enough to consider a position independent mass term. The RG equation is
defined as: 
\begin{equation}
\exp \left( -{\frac{1}{\hbar }}S_{k-\Delta k}(x)\right) =\int \mathcal{D}
\widetilde{x}\exp \left( -\frac{1}{\hbar }S_{k}(x+\widetilde{x})\right) \,.
\end{equation}
with the background path 
\begin{equation}
x(t)=x_{0}+\frac{1}{\beta }e^{iqt}x(q)+h.c.,
\end{equation}
and the fast fluctuating path 
\begin{equation}
\widetilde{x}=\int_{k}^{k-\Delta k}\frac{dp}{2\pi }e^{ipt}x(p)+h.c.
\end{equation}
We arrive at the following equation, giving the flow of $Z_{k}$ and $
U_{k}(x_{0})$ \cite{branchina}: 
\begin{equation}
Z_{k-\Delta k}q^{2}+U_{k-\Delta
k}^{(2)}(x_{0})=Z_{k}q^{2}+U_{k}^{(2)}(x_{0})+U_{k}^{(4)}(x_{0})\int_{k}^{k-
\Delta k}\frac{dp}{2\pi }\frac{1}{G(p)}+F(q)
\end{equation}
with 
\begin{equation}
F(q)=\frac{(U_{k}^{(3)})^{2}}{4}\int_{k}^{k-\Delta k}\frac{dp}{2\pi }
\int_{k}^{k-\Delta k}\frac{dp^{\prime }}{2\pi }\frac{\delta (p+p^{\prime
}+q)+\delta (p+p^{\prime }-q)}{G(p)G(p^{\prime })}+h.c.
\end{equation}
It's clear that for $q\leq \Delta k$ this integral gives contribution of the
form $\Delta k-\left| q\right| $ because the domain of integration is
deformed by the Dirac delta constraints. By expanding the denominator in
power of $q$ we get: 
\begin{equation}
F(q)=\frac{(U_{k}^{(3)})^{2}}{2}\frac{\Delta k-\left| q\right| }{G(k)G(k)}
(1+O(q^{2}))\text{ \qquad for }q\leq \Delta k
\end{equation}
and 
\begin{equation}
F(q)=0\text{ \qquad for }q\geq \Delta k.
\end{equation}
We see the generation of non-analytic terms which plague the computation
with a sharp cutoff. But as we have seen in the preceding section, the
integration mode after mode which is an exact computation gives $F(q)=0$.
So, the generation of non-analytical terms is in fact only an artifact of
the computation in the continuum and disappear if we make a rigorous
discrete calculus. The right result $F(q)=0$ can be recover in the continuum
if we suppose that $q>\Delta k$, because for that value of $q$, $F(q)=0$ in
the limit $\Delta k\rightarrow 0$. In this case, we face the problem already
encountered of the inconsistency between the equation of the potential and
its second derivative. But we have seen how to solve this problem in a very
natural way with the generalized potential.

In dimension higher than one we always encounter troubles with the sharp
cutoff formalism \cite{morris}\cite{wetterich}. It would be natural to
include a generalized potential, but terms like $F(q)$ are not necessary
equal to zero, because in field theory the domain of integration is
connected whereas in quantum mechanics it is disconnected, and this may be
the reason for the constant flow of $Z_{k}$. Equation (\ref{potcontinue1})
is the one dimensional version of the Wegner Houghton equation whereas (\ref
{eqZ0c1}) is very different from the equations obtained in field theory, for
example in references \cite{branchina}\cite{branchina1}. In those paper the
non-analytical terms where simply neglected and a non-perturbative equation
for $Z_{k}$ was proposed by considering the case $q\ll \Delta k$. This way
to treat the non-analytical terms has been already suggested in \cite
{wilson1} and \cite{sak}, but in our opinion no authors gave a convincing
arguments. The case $q\ll \Delta k$ yields then (neglected the
non-analytical terms) to: 
\[
Z_{k-\Delta k}=Z_{k}+\frac{(U_{k}^{(3)})^{2}}{2}\frac{\Delta k}{G(k)G(k)} 
\]
The motivation to consider this result as correct is that the equations for $
U_{k}(x_{0})$ and $U_{k}^{(2)}(x_{0})$ are now consistent. But, firstly it
is mathematically meaningless to neglect the non-analytic terms \cite{morris}
. Secondly, it is the merit of our computation to show without ambiguities
that this reasoning is incorrect as the discrete calculus leads to $
Z_{k-\Delta k}=Z_{k}.$

\section{Conclusion}

We have computed the renormalization group equation for the running action
in the context of a one quantum particle system. We have obtained this
equation for an arbitrary background configuration by introducing a
generalized action. This one is necessary to get a closed form for the
action under the renormalization group flow. This action contains much more
coupling constants than usual, and generates new flow equations for the
coupling constants. An important point is that our construction allows to
avoid some inconsistencies arising in the continuous formulation. We plan to
extend the generalized action construction in Field Theory.

\end{document}